\documentclass[pre,twocolumn]{revtex4}
\usepackage{epsfig}
\usepackage{bm}
\usepackage{amsmath}

\begin{document}

\title{Cross-helicity and transition from magnetohydrodynamics to kinetics in space plasmas}

\author{A. Bershadskii}

\affiliation{
ICAR, P.O. Box 31155, Jerusalem 91000, Israel
}

\begin{abstract}

 Cross-helicity dominated distributed chaos (CHDC) for magnetohydrodynamic (large-scale) and kinetic (small-scale) ranges of space plasmas has been studied using results of direct numerical simulations and of measurements made in the Earth's magnetosheath and in free solar wind by the instruments onboard of the Magnetospheric Multiscale Mission and CLUSTER spacecrafts. Existence of two generic types of transition from the magnetohydrodynamic (MHD) to kinetic range has been shown: 1) from the Kolmogorov-like (scaling) MHD turbulence to the kinetic CHDC and 2) from the magnetohydrodynamic CHDC to the kinetic CHDC. Results of measurements in the magnetosheath of some other planets: Saturn (Cassini observations), Jupiter (Galileo observations), and Mercury (Messenger observations), have been briefly discussed in this contest.

\end{abstract}

\maketitle

\section{Introduction}

 It is commonly excepted point of view that many important phenomena in the space plasma are governed by two ideal invariants: energy and cross-helicity (see, for instance, Refs. \cite{mgm}-\cite{ber1} and references therein). The cross-helicity can be considered as a measure of contribution of the Alfv\'enic waves. The existence of the two invariants greatly complicate understanding of the turbulent dynamics in both non-ionized fluids and plasmas, especially at the transition from the magnetohydrodynamic to the kinetic scales  \cite{matt3},\cite{ab},\cite{mil}. \\

   At sufficiently large scales, where the plasma can be considered as moving approximately as a single fluid (without macroscopic drift between the electrons and ions), the magnetohydrodunamic (MHD) description can be applied \cite{bc}. However, for the scales comparable to the electron and ion inertial lengths ($d_e$ and $d_i$) or to the Larmor radius ($\rho_e$ and $\rho_i$) one can expect appearance of some kinetic effects (although the plasma is still moving as a collective multifluid) \cite{sch}. The cross-helicty can be generalized from the MHD description on the kinetic one (at sufficiently large scales it becomes the MHD cross-helicty), and the kinetic cross-helicity is an ideal invariant together with the total energy \cite{mil}. \\
   
 It will be shown in Section III that in the both MHD (large-scale) and kinetic (small-scale) ranges the plasma dynamics can be described using notion of the cross-helicity dominated distributed chaos in the frames of the Kolmogorov-Iroshnikov phenomenology (cf the Ref. \cite{ber1}). This conclusion is supported by results of direct numerical simulations in Section IV.\\
 
   In Section V these results will be compared with results of measurements made in the Earth's magnetosheath and in free solar wind by the instruments onboard of the Magnetospheric Multiscale Mission and CLUSTER spacecrafts. Existence of two generic types of transition from the magnetohydrodynamic (MHD) to kinetic range will be also shown: 1) from the Kolmogorov-like (scaling) MHD turbulence to the kinetic CHDC and 2) from the magnetohydrodynamic CHDC to the kinetic CHDC.  Results of measurements in the magnetosheath of some other planets: Saturn (CASSINI observations \cite{had}), Jupiter (GALILEO observations \cite{tao}), and Mercury (MESSENGER observations \cite{huang}), have been briefly discussed in this contest.

\section{Energy and cross-helicity}

 The magnetohydrodynamic equations can be written in the Alfv\'enic units as
$$
 \frac{\partial {\bf v}}{\partial t} = - {\bf v} \cdot \nabla {\bf v} 
    -\frac{1}{\rho} \nabla {\cal P} - [{\bf b} \times (\nabla \times {\bf b})] + \nu \nabla^2  {\bf v} + {\bf f_v} \eqno{(1)}
$$
$$
\frac{\partial {\bf b}}{\partial t} = \nabla \times ( {\bf v} \times
    {\bf b}) +\eta \nabla^2 {\bf b} +  {\bf f_b} \eqno{(2)} 
$$
$$ 
\nabla \cdot {\bf v}=0, ~~~~~~~~~~~\nabla \cdot {\bf b}=0,  \eqno{(3,4)}
$$
where ${\bf v}$ and ${\bf b} = {\bf B}/\sqrt{\mu_0\rho}$  are the velocity and normalized magnetic field and they have the same dimension, $ {\bf f}_v$ and ${\bf f}_b$ are the forcing functions. \\

  In the non-dissipative ($\nu = \eta = 0$) and non-forcing ( ${\bf f_v} = {\bf f_b} =0$) case these equations have generally two quadratic invariants: the total energy
 $$
 {\mathcal E} = \int [{\bf v}({\bf x},t)^2 + {\bf b}({\bf x},t)^2 ] d{\bf r} \eqno{(5)} 
$$ 
 and cross-helicity
 $$
H_{cr} = \int {\bf v}({\bf x},t)\cdot {\bf B}({\bf x},t) d{\bf r}  \eqno{(6)}
$$

  When the mean magnetic field is zero there is an additional invariant - the magnetic helicity. For the inertial range of scales (if exists) these invariants can be considered as the adiabatic ones.\\
  
   In the Ref. \cite{mil} a simplified set of the model equations describing the low beta non-relativistic pair plasmas and the ultralow beta limit electron-ion plasmas in an imposed strong magnetic field $B_0\bf{\hat z}$ ($\bf{\hat z}$ is a unit vector along the axis ${\bf z}$) have been considered
$$
\frac{\partial }{\partial t}  \nabla^2_{\perp}\phi + \{\phi , \nabla^2_{\perp} \phi \}  = \{\psi , \nabla^2_{\perp}\psi \} + V_A \frac{\partial}{\partial z} \nabla^2_{\perp}\psi + f_{\phi} \eqno{(7)} 
$$
$$
 \frac{\partial }{\partial t}    (1-d_e^2\nabla^2_{\perp}) \psi + \{\phi , (1-d_e^2\nabla^2_{\perp})\psi \}  =  V_A \frac{\partial \phi}{\partial z} + f_{\psi} \eqno{(8)}
$$
Where the stream function $\phi$ is defined from ${\bf v_{\perp}}={\bf \hat{z}} \times {\boldsymbol \nabla}_{\perp} \phi$, and $\psi$ denotes the flux function for the perpendicular component of the magnetic field  $\delta {\bf B_{\perp}}/\sqrt{4 \pi \rho}={\bf \hat{z}} \times {\boldsymbol \nabla}_{\perp} \psi$, ${\bf B} = B_0{\bf \hat z} + \delta {\bf B_{\perp}}$ (and $\delta B_{\perp}/ B_0 \ll 1$), $V_{A}=B_{0} / \sqrt{4 \pi \rho}$ is the Alfv\'en speed,  $\{A,B\}= \partial_x A \partial_y B - \partial_x B \partial_y A$ is the Poisson bracket. \\

  These equations take into account the kinetic effects related to the electron inertia and characterized by the electron skin-depth $d_e$. They also have two quadratic invariants: the total energy 
$$
 \mathcal{E}=\frac{1}{2} \int \left\{\left(\nabla_{\perp} \psi\right)^{2}+d_{e}^{2}\left(\nabla_{\perp}^{2} \psi\right)^{2}+\left(\nabla_{\perp} \phi\right)^{2}\right\} d{\bf r} \eqno{(9)}
$$
and generalized cross-helicity
$$
{\mathcal H}=\int \left\{\nabla_{\perp}^{2} \phi\left(1-d_{e}^{2} \nabla_{\perp}^{2}\right) \psi\right\} d{\bf r}  \eqno{(10)}
$$

 It can be shown that ${\mathcal H} \simeq H_{cr}$ at the MHD scales $k_{\perp} d_e \ll 1$ ($k$ is the wavenumber) \cite{mil}. We will return to these equations below.\\
 %%%%%%%%%%%%%%% 1 %%%%%%%%%%%%%%%%%%
\begin{figure} \vspace{-1.65cm}\centering
\epsfig{width=.45\textwidth,file=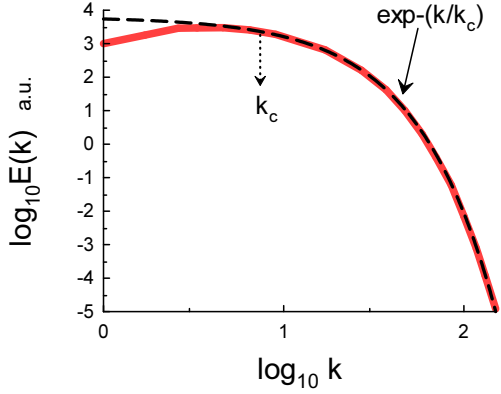} \vspace{-3.8cm}
\caption{ Magnetic energy spectrum for the DNS with {\it deterministic} forcing. } 
\end{figure}
%%%%%%%%%%%%%%%%%%%%%%%%%%%%%%%%%%%  

 %%%%%%%%%%%%%%% 2 %%%%%%%%%%%%%%%%%%
\begin{figure} \vspace{-1.2cm}\centering
\epsfig{width=.45\textwidth,file=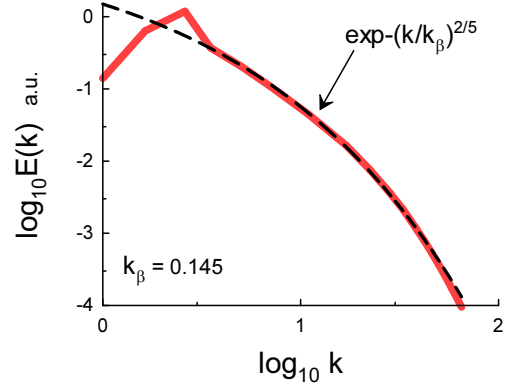} \vspace{-4.27cm}
\caption{Magnetic energy spectrum for the MHD DNS with a {\it random} helical forcing.} 
\end{figure}
%%%%%%%%%%%%%%%%%%%%%%%%%%%%%%%%%%%   

\section{Distributed chaos}

   At the onset of turbulence deterministic chaos is often characterized by the exponential power spectra \cite{mm},\cite{kds}
$$
E(k) \propto \exp-(k/k_c)  \eqno{(11)}
$$       
where $k$ is wavenumber. Figure 1, for instance, shows (in the log-log scales) magnetic energy spectrum obtained in recent direct numerical simulations (DNS) of the isotropic homogeneous MHD turbulence (the Eqs. (1-4)) with a large-scale {\it deterministic} forcing ${\bf f_v}$ (${\bf f_b}=0$). The spectral data used for the Fig. 1 were taken from Fig. 3 of the Ref. \cite{step}. The dashed curve indicates correspondence to the exponential spectrum Eq. (11) and the dotted arrow indicates position of the characteristic scale $k_c$. \\

      For a random forcing, however, the characteristic scale $k_c$ in the Eq. (11) fluctuates and to calculate the power spectra one needs in an ensemble averaging
$$
E(k) \propto \int_0^{\infty} P(k_c) \exp -(k/k_c)dk_c \propto \exp-(k/k_{\beta})^{\beta} \eqno{(12)}
$$    
with a probability distribution $P(k_c)$ characterizing the fluctuations. The stretched exponential power spectrum in the Eq. (12) is a natural generalization of the exponential power spectrum Eq. (11). \\

  The probability distribution $P(k_c)$ can be estimated from the Eq. (12) for large $k_c$  \cite{jon}
$$
P(k_c) \propto k_c^{-1 + \beta/[2(1-\beta)]}~\exp(-\gamma k_c^{\beta/(1-\beta)}) \eqno{(13)}
$$     
  
  Let us also consider a scaling relationship between characteristic fluctuations of magnetic field $B_c$
 and the characteristic scale  $k_c$
$$
B_c \propto  k_c^{\alpha}   \eqno{(14)}
$$

For the normally distributed $B_c$ a relationship between the exponents $\beta$ and $\alpha$
$$
\beta = \frac{2\alpha}{1+2\alpha}  \eqno{(15)}
$$
can be easily obtained from the Eqs. (13) and (14). \\

  Now we can use the dimensional considerations to obtain value of $\alpha$.  Namely
$$
B_c \propto    |\langle  {\bf v}\cdot {\bf B} \rangle|~\varepsilon^{-1/3}~ k_c^{1/3} \eqno{(16)} 
$$
here $\varepsilon$ is the total energy dissipation rate. The $\varepsilon$ and the average cross-helicity $h_{cr} =  \langle  {\bf v}\cdot {\bf B} \rangle$ can be considered as adiabatic invariants in the inertial range of scales (the Kolmogorov phenomenology). For $\alpha = 1/3$ one obtains from the Eq. (15) $\beta = 2/5$, i.e. the magnetic power spectrum
$$
E(k) \propto \exp-(k/k_{\beta})^{2/5}   \eqno{(17)}
$$  

  For the generalized (kinetic) cross-helicity the considerations are the same and corresponding magnetic energy spectrum is also the same Eq. (17).\\
  
     In the DNS reported in the papers Refs. \cite{tit},\cite{mag} the deterministic forcing used to obtain the spectrum shown in the Fig. 1 was replaced by a {\it random} forcing (with a controlled level of cross-helicity). The forcing was applied in the wavenumber range $1 \leq k \leq 3$ in the DNS terms.

   Figures 2 shows the spectrum of magnetic energy obtained in these DNS. The spectral data were taken from the Fig. 2b ($C=0.6$) of the Ref. \cite{tit}. The Reynolds and magnetic Reynolds numbers in these DNS were $Re =Re_m = 2094$. The dashed curve indicates correspondence to the stretched exponential spectrum Eq. (17).\\

%%%%%%%%%%%%%%%3 %%%%%%%%%%%%%%%%%%
\begin{figure} \vspace{-1.1cm}\centering
\epsfig{width=.47\textwidth,file=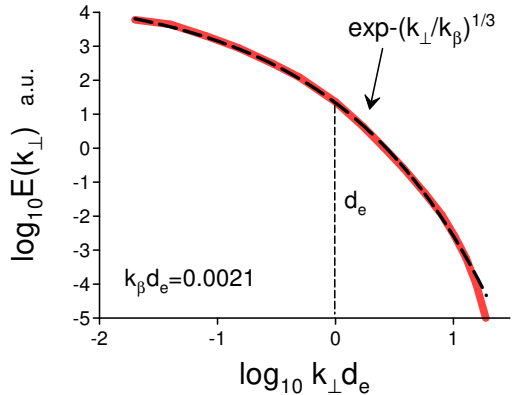} \vspace{-3.9cm}
\caption{Magnetic energy spectrum for the kinetic DNS (Eqs. 7-8) with a {\it random} helical forcing.} 
\end{figure}
%%%%%%%%%%%%%%%%%%%%%%%%%%%%%%%%%%%
%%%%%%%%%%%%%%% 4 %%%%%%%%%%%%%%%%%%
\begin{figure} \vspace{-1.2cm}\centering
\epsfig{width=.47\textwidth,file=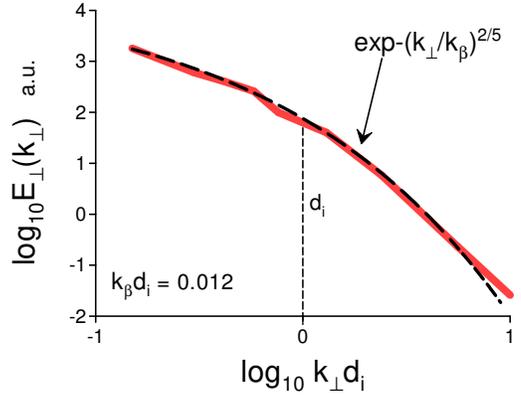} \vspace{-3.8cm}
\caption{Magnetic energy spectrum for the three-dimensional Vlasov-Maxwell system of equations numericaly solved using the particle-in-cell algorithm.} 
\end{figure}
%%%%%%%%%%%%%%%%%%%%%%%%%%%%%%%%%%%
%%%%%%%%%%%%%%% 5 %%%%%%%%%%%%%%%%%%
\begin{figure} \vspace{-0.5cm}\centering
\epsfig{width=.47\textwidth,file=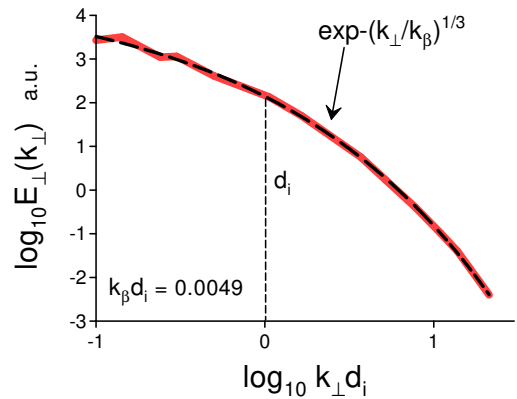} \vspace{-3.9cm}
\caption{Magnetic energy spectrum obtained in the three-dimensional hybrid particle-in-cell (CAMELIA) numerical simulation.}
\end{figure}
%%%%%%%%%%%%%%%%%%%%%%%%%%%%%%%%%%%

    A strong mean magnetic field can result in significant changes. The dimensional parameter dominating the inertial range of scales in the Kolmogorov phenomenology - $\varepsilon$, should be replaced by another parameter $\varepsilon b_0$ (where  $b_0 = B_0/\sqrt{\mu_0\rho}$ is the the normalized mean magnetic field) \cite{ir},\cite{kr}. \\
 
  With this replacement the Eq. (16) should be replaced by the equation 
$$
B_c \propto    |\langle  {\bf v}\cdot {\bf B} \rangle|~(\varepsilon b_0)^{-1/4}~ k_c^{1/4} \eqno{(18)} 
$$
i.e. $\alpha =1/4$ in this case and, consequently, we obtain from the  Eq. (15) $\beta = 1/3$, and the magnetic energy spectrum
$$
E(k) \propto \exp-(k/k_{\beta})^{1/3}   \eqno{(19)}
$$  

 For the generalized (kinetic) cross-helicity the considerations are the same and corresponding magnetic energy spectrum is also the same Eq. (19).\\
 
 \section{Numerical simulations of kinetic plasma}
 
   Figure 3 shows magnetic energy spectrum obtained in a direct numerical simulation based on the the Eqs. (7-8) in a triply periodic spatial domain. The energy forcing was random and delta-correlated. The number ${\mathcal R_H} = {\mathcal H}_{inj}^+/{\mathcal H}_{inj}^-$ is a measure of the disbalance between positive and negative generalized (kinetic) cross-helicity injection. In this case ${\mathcal R_H} =10$. The spectral data were taken from Fig. 1a of the Ref. \cite{mil}. The dashed curve indicates correspondence to the stretched exponential spectrum Eq. (19) and the vertical dashed line indicates position of the electron inertial length $d_e$. \\
%%%%%%%%%%%%%%% 6 %%%%%%%%%%%%%%%%%%
\begin{figure} \vspace{-0.9cm}\centering
\epsfig{width=.47\textwidth,file=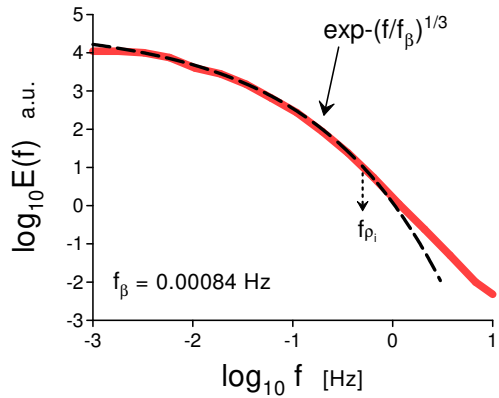} \vspace{-4.3cm}
\caption{Trace power spectrum of magnetic field measured at the Cluster spacecraft mission in the Earth's magnetosheath. }
\end{figure}
%%%%%%%%%%%%%%%%%%%%%%%%%%%%%%%%%%%   
 %%%%%%%%%%%%%%% 7 %%%%%%%%%%%%%%%%%%
\begin{figure} \vspace{-0.5cm}\centering
\epsfig{width=.45\textwidth,file=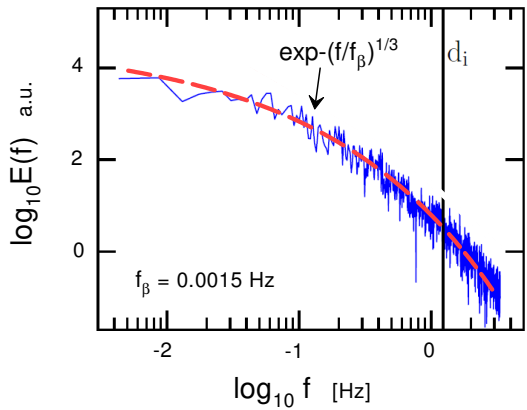} \vspace{-3.25cm}
\caption{Magnetic energy spectrum in the Earth's magnetosheath (Magnetospheric Multiscale Mission spacecraft).}
\end{figure}
%%%%%%%%%%%%%%%%%%%%%%%%%%%%%%%%%%%
%%%%%%%%%%%%%%% 8 %%%%%%%%%%%%%%%%%%
\begin{figure} \vspace{-1.55cm}\centering
\epsfig{width=.47\textwidth,file=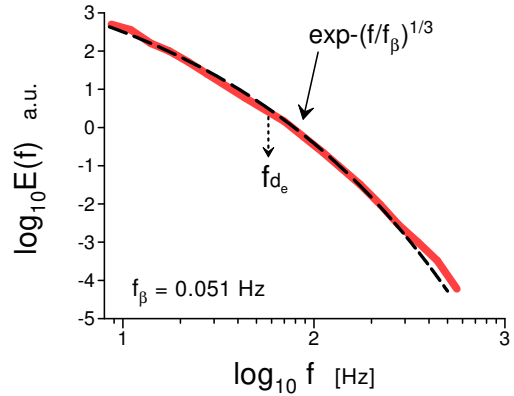} \vspace{-3.675cm}
\caption{Trace power spectrum of magnetic field measured at the Cluster spacecraft mission in the Earth's magnetosheath kinetic range of scales (the STAFF-SA experiment). }
\end{figure}
%%%%%%%%%%%%%%%%%%%%%%%%%%%%%%%%%%% 

  It should be noted that the kinetic numerical simulations of the plasma with imposed magnetic field not always show the spectrum Eq. (19) but also the spectrum Eq. (17). Figure 4, for instance, shows magnetic energy spectrum obtained in a fully kinetic three dimensional simulation of collisionless plasma turbulence (the spectral data were taken from Fig. 2 of Ref. \cite{wan}). In this simulation the three-dimensional Vlasov-Maxwell system of equations was numericaly solved using the particle-in-cell (PIC) algorithm. A uniform plasma state with Maxwellian-distributed electrons and ions of equal temperature was taken as initial condition. The total number of particles was 1.7 trillion. The initial large-scale `turbulent' spectrum for magnetic fluctuations had polarization transverse to the imposed mean magnetic field. The statistical noise was removed by a low pass filter. Appearance of highly structured sheet-like current density structures, spanning through the range of scales between the electron and proton inertial scales, was observed in this numerical simulation. 
  
  The dashed curve indicates correspondence to the stretched exponential spectrum Eq. (17) and the vertical dashed line indicates position of the ion inertial length $d_i$.\\
    
  Figure 5 shows magnetic energy spectrum obtained in a recent three dimensional simulation of collisionless plasma turbulence using a hybrid particle-in-cell CAMELIA code (the spectral data were taken from Fig. 1d of Ref. \cite{cgf}).  In this simulation the equliribrium plasma with uniform density and temperature was perturbed with ion bulk velocity and magnetic field random fluctuations, perpendicular to the imposed magnetic field. The data shown in the Fig. 5 were obtained at at the peak of the turbulent activity. In the  hybrid particle-in-cell CAMELIA code the ions are considered as macroparticles whereas the massless electrons are considered as a charge neutralizing fluids. The boundary conditions were periodic in the three dimensions. The dashed curve indicates correspondence to the stretched exponential spectrum Eq. (19) and the vertical dashed line indicates position of the ion inertial length $d_i$.

\section{Space plasmas}

\subsection{Earth's magnetosheath}

  In a recent paper Ref. \cite{teo} the magnetic energy spectrum was calculated using the data obtained by the Cluster Fluxgate Magnetometer onboard the Cluster spacecraft operated in the Earth's magnetosheath under slow solar wind conditions. The magnetosheath magnetic field was, for the most part of the measurements, perpendicular to the mean flow direction. 
  
  Figure 6 shows the trace of the spectral matrix vs frequency. The spectral data were taken from Fig. 5a of the Ref. \cite{teo}. For this case the mean velocity of the plasma $|\langle {\bf V} \rangle|$ (in the spacecraft frame) was significantly larger than the representative fluctuations of the velocity. Therefore, the Taylor "frozen" turbulence hypothesis can be applied \cite{hb},\cite{tbn}. It suggests that the temporal dynamics of the fields measured by the probe reflects spatial structures convected past the probe by the mean velocity. Therefore, it is not a truly frequency spectrum but a reflection of a wavenumber spectrum with the replacement  $k \simeq 2\pi f/ |\langle {\bf V}| \rangle$ . Hence, the dashed curve in the Fig. 6 indicates the {\it wavenumber} spectrum Eq. (19) and the scale $k_{\beta}\simeq 2\pi f_{\beta}/ |\langle {\bf V} \rangle|$ . The vertical dashed arrow indicates position of the ion Larmor radius $\rho_i$.\\

  In another recent paper Ref. \cite{ban} the magnetic energy spectrum was calculated using the data obtained by Magnetospheric Multiscale Mission spacecraft operated in the Earth's magnetosheath \cite{bur}. The low density fluctuations and the low Mach number (the Table 1 of the Ref. \cite{ban}) can justify applicability of the incompressible MHD for the large scales. The normalized cross-helicity $\sigma_c \simeq 0.24$ (the Table 2 of the Ref. \cite{ban}) was smaller than the ones observed in the free solar wind but it was rather not negligible (the maximal value of $\sigma_c$ is 1).\\
  
    Figure 7 shows the frequency spectrum of magnetic energy (the spectral data were taken from the Fig. 2 of the Ref. \cite{ban}). For this case the mean velocity of the plasma $|\langle {\bf V} \rangle|$ (in the spacecraft frame) also was significantly larger than the representative fluctuations of the velocity. Therefore, the Taylor "frozen" turbulence hypothesis can be applied in this case as well. Hence, the dashed curve in the Fig. 7 indicates the {\it wavenumber} spectrum Eq. (19). The vertical black bar indicates the ion inertial length $d_i$.\\
    
    To obtain information about the scales comparable to the electron inertial lenght ($d_e$) one needs data measured at scales smaller than those shown in Figs. 6-7. Figure 8 shows a trace power spectrum of magnetic field measured at the Cluster spacecraft mission (the STAFF-SA experiment \cite{corn}) in the Earth's magnetosheath. The spectral data were taken from the Fig. 1a of the Ref. \cite{macl}. To obtain the spectra more relevant to the background turbulence at small scales the authors of the Ref. \cite{macl} removed contribution from the whistlers (local wave activity typically observed at the electron scales). They also have shown applicability of the Taylor hypothesis for the spectra filtered from the whistlers. Therefore, the dashed curve in the Fig. 8 indicates the {\it wavenumber} spectrum Eq. (19). The mean magnetic field $B_0 \simeq 40 nT$ in this case. The frequency $f_{de}$ corresponds to the electron inertial length. \\

  %%%%%%%%%%%%%%% 9 %%%%%%%%%%%%%%%%%%
\begin{figure} \vspace{-1.5cm}\centering
\epsfig{width=.47\textwidth,file=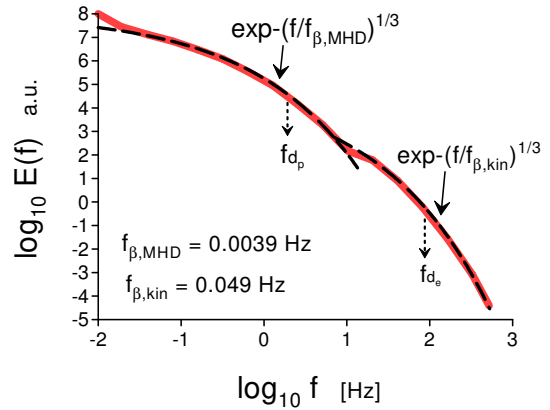} \vspace{-3.9cm}
\caption{Trace power spectrum of magnetic field measured at the Cluster spacecraft mission in the Earth's magnetosheath (MHD scales -  the FGM/EFW instrument, and the kinetic scales - the STAFF-SA instrument). }
\end{figure}
%%%%%%%%%%%%%%%%%%%%%%%%%%%%%%%%%%%
%%%%%%%%%%%%%%% 10 %%%%%%%%%%%%%%%%%%
\begin{figure} \vspace{-0.5cm}\centering
\epsfig{width=.47\textwidth,file=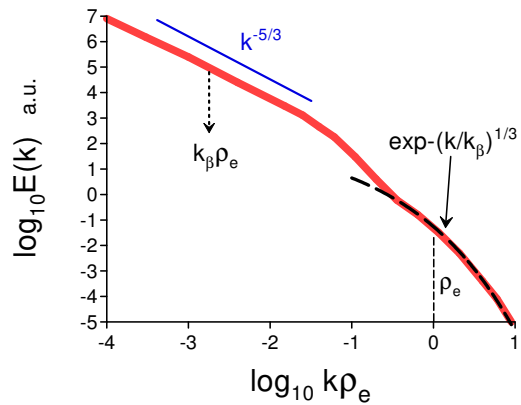} \vspace{-3.7cm}
\caption{Normalized wavenumber magnetic power spectrum (averaged over 7 time periods) for free solar wind at the Earth's orbit (at 1 AU). }
\end{figure}
%%%%%%%%%%%%%%%%%%%%%%%%%%%%%%%%%%%
%%%%%%%%%%%%%%% 11 %%%%%%%%%%%%%%%%%%
\begin{figure} \vspace{-1.5cm}\centering
\epsfig{width=.47\textwidth,file=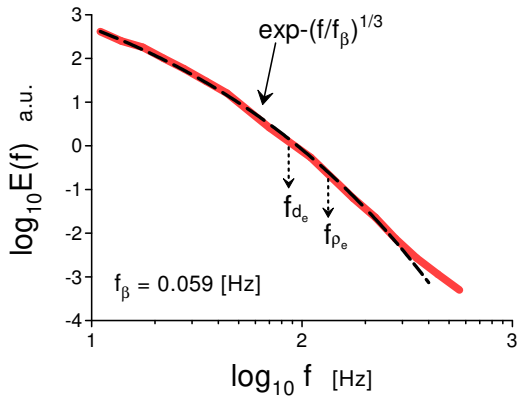} \vspace{-3.9cm}
\caption{Magnetic power spectrum for the free solar wind in the kinetic range (with the noise filtered out).}
\end{figure}
%%%%%%%%%%%%%%%%%%%%%%%%%%%%%%%%%%%      

    Figure 9 shows a trace power spectrum of magnetic field measured at the Cluster spacecraft mission (for the MHD scales the data were obtained using the FGM/EFW instrument and for the kinetic range the data were obtained using the STAFF-SA instrument, cf Fig. 8). The spectral data were taken from the Fig. 4a of the Ref. \cite{macl}. The frequency $f_{dp}$ corresponds to the proton inertial length. The mean magnetic field $B_0 \simeq 24 nT$ in this case. The dashed curve in the Fig. 9 indicates the {\it wavenumber} spectrum Eq. (19).\\
    
\subsection{Free solar wind at the Earth's orbit}    

  In the Refs. \cite{ale1},\cite{ale2} measurements of the Cluster instruments were also used for investigation of the free solar wind. 
  
  Figure 10 shows a normalized wavenumber magnetic power spectrum (averaged over 7 time periods) in free solar wind (slow and fast) at the Earth's orbit (at 1 AU). The wavenumber spectrum was obtained using the Taylor hypothesis and the wavenumbers $k$ were normalized by the electron Larmor radius ($\rho_e$). The spectral data were taken from Fig. 4c of the Ref. \cite{ale1}. The vertical dashed line indicates position of the electron Larmor radius. The dashed curve in the Fig. 10 indicates the  wavenumber spectrum Eq. (19) for the small-scale (kinetic) region and the straight solid line indicates the Kolmogorov-like spectrum $E(k) \propto k^{-5/3}$ for the large-scale (MHD) region.  \\
  
    Figure 11 shows a frequency magnetic power spectrum in the free solar wind in the small-scale (kinetic) range with the STAFF-SA instrument's noise filtered out. The spectral data were taken from Fig. 2a of the Ref. \cite{ale2} (cf Fig. 8 for the Earth's magnetosheath). The dashed curve in the Fig. 11 indicates the  stretched exponential spectrum Eq. (19). \\

    The spectra shown in the Figs. 6-11 were interpreted using the Taylor's hypothesis. In a recent paper Ref. \cite{rne} results of the spectral measurements performed directly in the wave-vector 3D-space (without invoking the Taylor’s hypothesis) were reported.  The four-point magnetometer data (obtained at the Cluster spacecraft mission in the solar wind) were analyzed using an interferometric wave-vector analysis. 
    
    Figures 12 and 13 show the magnetic power spectra as function of $k_{\perp}$ and $k_{\parallel}$ correspondingly. The spectral data were taken from Figs. 2a and 2b of the Ref. \cite{rne}. The dashed curves in the Figs. 12 and 13 indicate the stretched exponential spectrum Eq. (19) and the vertical dashed lines indicate the position of the proton inertial length.  
  
 %%%%%%%%%%%%%%% 12 %%%%%%%%%%%%%%%%%%
\begin{figure} \vspace{-1.5cm}\centering
\epsfig{width=.47\textwidth,file=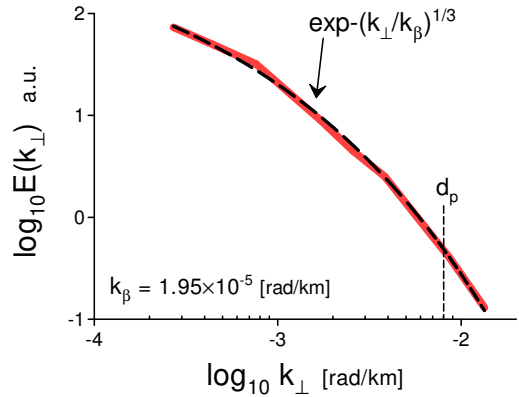} \vspace{-3.9cm}
\caption{Magnetic power spectrum vs $k_{\perp}$ for the free solar wind.}
\end{figure}
%%%%%%%%%%%%%%%%%%%%%%%%%%%%%%%%%%%     
%%%%%%%%%%%%%%% 13 %%%%%%%%%%%%%%%%%%
\begin{figure} \vspace{-0.5cm}\centering
\epsfig{width=.47\textwidth,file=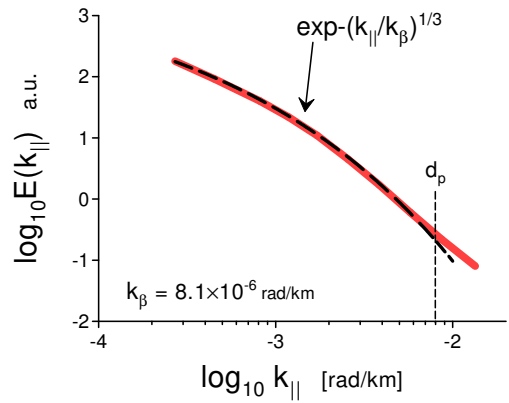} \vspace{-3.9cm}
\caption{Magnetic power spectrum vs $k_{\parallel}$ for the free solar wind.}
\end{figure}
%%%%%%%%%%%%%%%%%%%%%%%%%%%%%%%%%%%    

 \subsection{Saturn's, Jupiter's, and Mercury's magnetosheath }   
 
 %%%%%%%%%%%%%%% 14 %%%%%%%%%%%%%%%%%%
\begin{figure} \vspace{-0.4cm}\centering
\epsfig{width=.48\textwidth,file=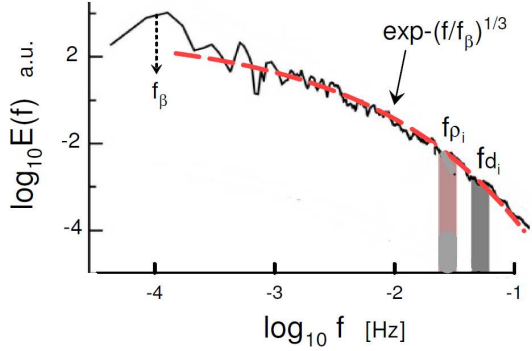} \vspace{-5.1cm}
\caption{Power spectrum of magnetic field fluctuations measured at the Cassini spacecraft mission in the Saturn's magnetosheath. }
\end{figure}
%%%%%%%%%%%%%%%%%%%%%%%%%%%%%%%%%%%
%%%%%%%%%%%%%%% 15 %%%%%%%%%%%%%%%%%%
\begin{figure} \vspace{-0.4cm}\centering
\epsfig{width=.47\textwidth,file=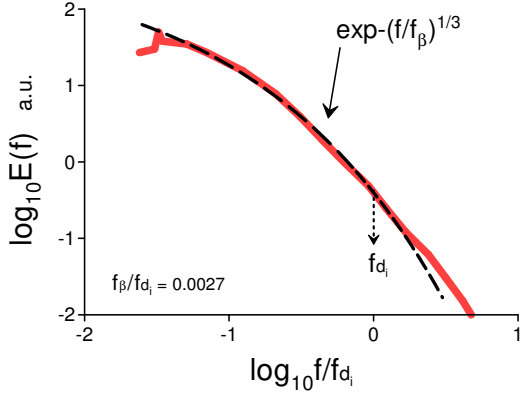} \vspace{-3.8cm}
\caption{Power spectrum of magnetic field fluctuations measured at the Galileo spacecraft mission in the Jupiter's magnetosheath.  }
\end{figure}
%%%%%%%%%%%%%%%%%%%%%%%%%%%%%%%%%%%
%%%%%%%%%%%%%%% 16 %%%%%%%%%%%%%%%%%%
\begin{figure} \vspace{-1.2cm}\centering
\epsfig{width=.47\textwidth,file=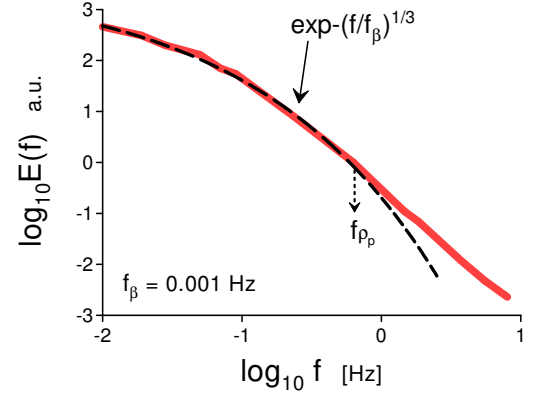} \vspace{-3.7cm}
\caption{Power spectrum of magnetic field fluctuations measured at the Messenger spacecraft mission in the Mercury magnetosphere.  }
\end{figure}
%%%%%%%%%%%%%%%%%%%%%%%%%%%%%%%%%%%

   It is interesting to compare the above obtained results with the results of measurements obtained in the magnetosheath of other planets. \\
   
   Figure 14 shows power spectrum of the magnetic field fluctuations computed using the data obtained by the Flux Gate Magnetometer (MAG) onboard of the Cassini spacecraft in the Saturn's magnetosheath (cf. Fig. 7). The spectral data were taken from Fig. 2 of the Ref. \cite{had}. The dashed curve in the Fig. 14 indicates the  stretched exponential spectrum Eq. (19). The position of the $f_{\beta}$ (the dotted arrow) indicates that the large-scale coherent structures (the large-scale peak) dominate the entire distributed chaos in this case. The two shaded bands indicate the ion Larmor radius and inertial length (the width of the bands reflects the corresponding uncertainty).\\
   
   Figure 15 shows total power spectrum of the magnetic field fluctuations computed using the data obtained by the MAG instrument onboard of the Galileo spacecraft in the Jupiter's magnetosheath. The spectral data were taken from Fig. 10d of the Ref. \cite{tao} and correspond to the radial distance bin $20 - 25R_j$ (the nearest to Jupiter), where $R_j$ is the radius of Jupiter. The authors of the Ref. \cite{tao} provides reasons for applicability of the Taylor hypothesis in this case. The dashed curve in the Fig. 15 indicates the  stretched exponential spectrum Eq. (19).\\
   
   Figures 16 and 17 show power spectra of the magnetic field fluctuations computed using the data obtained by the MAG instrument onboard of the Messenger spacecraft in the Mercury magnetosphere and magnetosheath (close to magnetopause) respectively. The spectral data were taken from Figs. 6d and 6c of the Ref. \cite{huang}. The frequency $f{\tiny \rho_p}$ corresponds to the proton Larmor radius. The dashed curve in the Figs. 16 and 17 indicates the  stretched exponential spectrum Eq. (19).\\

\section{Conclusions}

  Comparing the Figs. 6-9 one can conclude that in the Earth's magnetosheath plasma:
  
  A. For both MHD and kinetic scales regions magnetic energy spectra can be described by the stretched exponential spectrum Eq. (19) (the cross-helicty dominated distributed chaos in the frames of the Iroshnikov phenomenology). 
  
  B. The value of the parameter $k_{\beta}$ are significantly different for these two scale regions $k_{\beta, MHD} \ll k_{\beta, kin}$. 
  
  C. The transition between the MHD and kinetic distributed chaos regions takes place between the $d_i$ and $d_e$ scales.\\

%%%%%%%%%%%%%%% 17 %%%%%%%%%%%%%%%%%%
\begin{figure} \vspace{-1.5cm}\centering
\epsfig{width=.47\textwidth,file=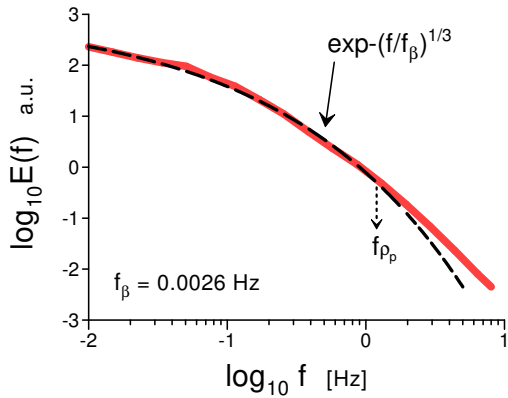} \vspace{-3.9cm}
\caption{Power spectrum of magnetic field fluctuations measured at the Messenger spacecraft mission in the Mercury magnetosheath (close to magnetopause).  }
\end{figure}
%%%%%%%%%%%%%%%%%%%%%%%%%%%%%%%%%%%

  Comparing the Figs. 8 and 11 one can conclude that in the kinetic scale region for the Earth's magnetosheath plasma and for the free solar wind plasma on the Earth's orbit (1 AU) the magnetic energy spectra are rather similar. However, in the case of the solar wind the transition from the MHD range of scales to the kinetic range can be a transition from the Kolmogorov-like MHD spectrum to the stretched exponential spectrum Eq. (19) characterizing the cross-helicty dominated distributed chaos (cf. Figs. 9 and 10). \\


\begin{thebibliography}{99}

\bibitem{mgm} W.H. Matthaeus, M.L. Goldstein, and D.C. Montgomery, Phys. Rev. Lett., {\bf 51}, 1484 (1983).
\bibitem{chb} M. Christensson, M. Hindmarsh and A. Brandenburg, Phys. Rev. E, {\bf 64}, 056405 (2001).
\bibitem{bg1} A. Briard and T. Gomez, J. Plasma Phys., {\bf 84}, 905840110 (2018).
\bibitem{pip}  V.V. Pipin and N. Yokoi, ApJ, {\bf 859}, 18 (2018).
\bibitem{zb}  H Zhang and A. Brandenburg, ApJ Letters, {\bf 862}, L17 (2018).
\bibitem{bo} A. Brandenburg S. Oughton,  Astron. Nachr. {\bf 339}, 631 (2018).
\bibitem{sorr} L. Sorriso-Valvo, F. Carbone, S. Perri,  A. Greco, R. Marino, R. Bruno, Solar Phys., {\bf 293}, 10 (2018)
\bibitem{pou1}  A. Pouquet, J.E. Stawarz, D. Rosenberg and R. Marino, Earth and Space Sciences, {\bf 6}, 351 (2019)
\bibitem{ber1} A. Bershadskii, Res. Notes AAS, {\bf 4}, 10 (2020).
\bibitem{matt3} W.H. Matthaeus, M. Wan, S. Servidio, A. Greco, K.T. Osman, S. Oughton and P. Dmitruk, Phil. Trans. Royal Soc. A, 373, 10.1098/rsta.2014.0154 (2015)
\bibitem{ab} A. Alexakis and L. Biferale, Phys. Rep., {\bf 767-769}, 1 (2018)
\bibitem{mil} L.M. Milanese, N.F. Loureiro, M. Daschner and S. Boldyrev, Phys. Rev. Lett., {\bf 125}, 265101 (2020)
\bibitem{had} L.Z. Hadid, F. Sahraoui, K.H. Kiyani, A. Retino, R. Modolo, P. Canu, A. Masters, and M. K. Dougherty, ApJ Lett., {\bf 813}, L29 (2015)
 \bibitem{tao} C. Tao, F. Sahraoui, D. Fontaine, J. de Patoul, T. Chust, S. Kasahara and A. Retino, J. Geophys. Res. Space Physics, {\bf 120}, 2477 (2015)
\bibitem{huang} S.Y. Huang, Q.Y. Wang, F. Sahraoui et al., ApJ, {\bf 891} 159 (2020)
\bibitem{bc} R. Bruno and V. Carbone, Living Rev. Sol. Phys., {\bf 2}(4), (2005)
\bibitem{sch} A.A. Schekochihin, S.C. Cowley,  W. Dorland, G.W. Hammett, G.G.Howes, E. Quataert and T. Tatsuno, Astrophys. J. Suppl. Ser., {\bf 182}, 310 (2009)
\bibitem{mm} J. E. Maggs and G. J. Morales, Phys. Rev. Lett. {\bf 107},
185003 (2011); Phys. Rev. E {\bf 86}, 015401(R) (2012); Plasma Phys. Control.
Fusion {\bf 54} 124041 (2012).
\bibitem{kds} S. Khurshid, D.A. Donzis, and K.R. Sreenivasan, Phys. Rev. Fluids, {\bf 3}, 082601(R) (2018).
\bibitem{step} R. Stepanov, A. Teimurazov, V. Titov, M.K. Verma, S. Barman, A. Kumar, A. and F. Plunian, in 'Ivannikov ISPRAS Open Conference (ISPRAS)' pp. 90-96 (2017).  https://ieeexplore.ieee.org/document/8273304
\bibitem{jon} D.C. Johnston, Phys. Rev. B, {\bf 74}, 184430 (2006).
\bibitem{tit} V.V. Titov, Computational Continuum Mechanics, {\bf 12}, 5 (2019).
\bibitem{mag} V.Titov, R. Stepanov, N.Yokoi, M.Verma and R. Samtaney, Magnetohydrodynamics, {\bf 55}, 225 (2019).
\bibitem{ir} R.S. Iroshnikov, Astronomicheskii Zhurnal, {\bf 40}, 742 (1963) (English translation in Soviet Astronomy
{\bf 7}, 566 (1964)).
\bibitem{kr} R.H. Kraichnan, Phys. Fluids, {\bf 8}, 1385 (1965)
\bibitem{wan} M. Wan, W.H. Matthaeus, V. Roytershteyn, H. Karimabadi, T. Parashar, P. Wu, and M. Shay,
Phys. Rev. Lett., {\bf 114}, 175002 (2015)
\bibitem{cgf} S.S. Cerri , D. Groselj and L. Franci, Frontiers in Astronomy and Space Sciences, {\bf 6}, 64 (2019)
\bibitem{teo} E. Teodorescu, M. Echim and G. Voitcu, ApJ, {\bf 910}, 66 (2021)
\bibitem{hb} T.S. Horbury and A. Balogh, Nonlin. Proc. Geophys., {\bf 4}, 185 (1997)
\bibitem{tbn} R.A. Treumann, W. Baumjohann and Y. Narita, Earth, Planets and Space, {\bf 71}, 41 (2019)
\bibitem{ban} R. Bandyopadhyay et al., ApJ, {\bf 866} 106 (2018).
\bibitem{bur} J.L Burch, T.E. Moore, R.B. Torbert and B.L. Giles, SSRv, {\bf 199}, 5 (2016)
\bibitem{corn} N. Cornilleau-Wehrlin N. et al., Space Sci. Rev., {\bf 79}, 107 (1997)
\bibitem{macl} L. Matteini, O. Alexandrova, C.H.K. Chen and C. Lacombe, MNRAS, {\bf 466}, 945 (2017)
\bibitem{ale1} O. Alexandrova, J. Saur, C. Lacombe, A. Mangeney, J. Mitchell, S. J. Schwartz and P. Robert, Phys. Rev. Lett., {\bf 103}, 165003 – (2009)
 \bibitem{ale2} O. Alexandrova, J. Saur, C. Lacombe, A. Mangeney, S. J. Schwartz, J. Mitchell, R. Grappin, and P. Robert, AIP Conference Proc., {\bf 1216}, 144 (2010)
 \bibitem{rne} O.W. Roberts, Y. Narita, and C.P. Escoubet, ApJ Lett., {\bf 851}, L11 (2017)




\end{thebibliography}
\end{document}